\documentclass[usenatbib,usegraphicx]{mn2e}
\usepackage{amsmath,amssymb,xspace,hyperref}

\newcommand{\SMILE}{\textsc{smile}\xspace}
\newcommand{\Raga}{\textsc{Raga}\xspace}
\newcommand{\phiGRAPE}{$\phi$\textsc{grape}\xspace}
\newcommand{\Nbody}{$N$-body\xspace}
\renewcommand{\d}{\partial}

\newcommand{\dvpar}{\langle \Delta v_\| \rangle}
\newcommand{\dvsqpar}{\langle \Delta v^2_\| \rangle}
\newcommand{\dvsqper}{\langle \Delta v^2_\bot \rangle}
\newcommand{\rmax}{r_\mathrm{max}}

\newcommand{\tdyn}{T_\mathrm{dyn}}
\newcommand{\trel}{T_\mathrm{rel}}
\newcommand{\trelhm}{T_\mathrm{rel,h-m}}
\newcommand{\rh}{r_\mathrm{m}}
\newcommand{\Lcirc}{L_\mathrm{circ}}

\title[Monte Carlo method for non-spherical systems]
{A new Monte Carlo method for dynamical evolution of non-spherical stellar systems}

\author[E. Vasiliev]{Eugene Vasiliev$^1$\thanks{E-mail: eugvas@lpi.ru}\\
$^1$Lebedev Physical Institute, Moscow, Russia }

\date{Accepted 2014 November 4.  Received 2014 November 3; in original form 2014 April 1}

\begin{document}
\volume{446} \pagerange{3150--3161} \pubyear{2015}

\maketitle

\label{firstpage}

\begin{abstract}
We have developed a novel Monte Carlo method for simulating the dynamical evolution 
of stellar systems in arbitrary geometry. 
The orbits of stars are followed in a smooth potential represented by a basis-set 
expansion and perturbed after each timestep using local velocity diffusion 
coefficients from the standard two-body relaxation theory. 
The potential and diffusion coefficients are updated after an interval of time that is 
a small fraction of the relaxation time, but may be longer than the dynamical time. 
Thus our approach is a bridge between the Spitzer's formulation of the Monte Carlo 
method and the temporally smoothed self-consistent field method. 
The primary advantages are the ability to follow the secular evolution of shape 
of the stellar system, and the possibility of scaling the amount of two-body 
relaxation to the necessary value, unrelated to the actual number of particles 
in the simulation. 
Possible future applications of this approach in galaxy dynamics include the problem 
of consumption of stars by a massive black hole in a non-spherical galactic nucleus, 
evolution of binary supermassive black holes, 
and the influence of chaos on the shape of galaxies, 
while for globular clusters it may be used for studying the influence of rotation.
\end{abstract}


\section{Introduction}

Many problems of stellar dynamics deal with self-gravitating systems which are 
in dynamical equilibrium, but slowly evolve due to two-body relaxation or some 
other factor, such as a massive black hole or the diffusion of chaotic orbits.
The most general method of studying these systems is a direct \Nbody simulation, 
however, in many cases it turns out to be too computationally expensive. 
Alternative methods, such as Fokker--Planck, gaseous, or Monte Carlo models, 
have historically been developed mostly for spherical star clusters. 
In this paper we present a formulation of the Monte Carlo method suitable for 
non-spherical stellar systems.

The paper is organized as follows. Section~\ref{sec:overview} reviews the existing 
simulation methods and outlines the motivation for the proposed new formulation;
Section~\ref{sec:relaxation} presents the theoretical background of two-body 
relaxation theory; Section~\ref{sec:raga} discusses the implementation of 
the non-spherical Monte Carlo code and Section~\ref{sec:tests} presents the results 
of test simulations. Section~\ref{sec:conclusions} lists possible applications of 
the new method and sums up.

\section{Overview of numerical methods}  \label{sec:overview}

The development of Monte Carlo methods for simulation of star clusters started in 
early 1970s with two different approaches, pioneered by Spitzer and H\'enon.

In the original formulation of \citet{SpitzerHart1971a}, the motion of test stars 
in a spherically symmetric potential was followed numerically on the dynamical 
timescale, and perturbations to the velocity was computed assuming a Maxwellian 
distribution of background stars (scatterers), with the mean density and velocity 
dispersion computed in 25 radial bins by averaging over 40 stars in each bin; 
thus, the test stars were also used for determining the smoothed properties of 
the field stars. To speed up computation, dependence of velocity diffusion 
coefficients on the velocity of the test star was ignored 
(the values corresponded to the average thermal velocity); 
this simplification was lifted in \citet{SpitzerThuan1972}. 
Since perturbations to each star's velocity are independent of each other, 
the global conservation of energy is not guaranteed; thus a correction is applied 
after each timestep which cancels the residual fluctuations.
This method became known as the ``Princeton'' Monte Carlo code \citep{Spitzer1975}.

In another variant of this method, \citet{SpitzerShapiro1972} turned to using 
the diffusion coefficients in energy $E$ and angular momentum $L$, averaged over 
the radial period of the test star. This approach was subsequently developed by 
\citet{ShapiroMarchant1978} to study the steady-state solution for 
the distribution of stars around a massive black hole: the potential was assumed 
to be dominated by the point mass, the diffusion coefficients in $E$ and $L$ were
computed self-consistently from the distribution function $f(E,L)$, which was 
then adjusted iteratively until convergence. The capture of low angular momentum 
stars by the black hole was also taken into account, which necessitated a rather 
complex scheme for choosing the timestep: it was determined by the relaxation time 
but also required not to miss a potentially disruptive periapsis passage near 
the black hole. It also had an ingenious scheme for particle cloning (mass 
refinement) to allow for better sampling of phase-space close to the black hole.
Subsequent papers extended the method to self-consistent (rather than 
point-mass-dominated) potentials \citep{MarchantShapiro1979} and to evolutionary 
simulation including the heating by the black hole, core collapse, and 
evaporation \citep{MarchantShapiro1980}.
This approach has been dubbed the ``Cornell'' code \citep{Shapiro1985}.
More recently, \citet{Hopman2009} and \citet{MadiganHL2011} have used this 
formulation to study the dynamics around massive black holes.

At the same time, \citet{Henon1971a,Henon1971b} introduced another variant of 
Monte Carlo method, in which pairs of stars are interacting directly 
\citep[see also][]{Henon1967}. 
Unlike the conventional \Nbody simulations, these pairwise interactions are 
computed only between particles that are adjacent in radius. For each pair of 
interacting particles, their relative velocity is changed by an amount which 
reproduces statistically the effect of many individual encounters during 
the same interval of time. 
The timestep is chosen to be a fraction of the  relaxation time $\trel$, instead 
of a fraction of the dynamical time $\tdyn$. After each timestep, the stars are 
assigned new positions (or, rather, radii, since the system is assumed to be 
spherically symmetric).
This method was subsequently improved by \citet{Stodolkiewicz1982}, who included
a variable timestep (proportional to the radius-dependent $\trel$), correction of 
velocities due to the changes in potential after recomputing new positions of 
particles, continuous stellar mass spectrum, and shock heating due to passages of 
the globular cluster through the galactic disc. 
\citet{Stodolkiewicz1986} introduced many other physical ingredients such as 
stellar evolution, primordial binaries \citep[also studied by][]{SpitzerMathieu1980}
and cross-sections for three- and four-body interactions, and stellar collisions.

All presently used codes follow the H\'enon's approach. Since late 1990s, two 
groups \citep{Giersz1998,JoshiRPZ2000} have been developing sophisticated codes 
including much additional physics beyond two-body relaxation: parametrized single 
and binary stellar evolution \citep{HurleyPT2000,HurleyTP2002,ChatterjeeFUR2010}, 
direct integration of few-body encounters \citep{GierszSpurzem2003,FregeauRasio2007},
accurate treatment of escapers \citep{FukushigeHeggie2000}. 
The present versions of these codes are described in \citet{GierszHHH2013} and 
\citet{Pattabiraman2013}. 
In these codes, the number of particles in the simulation equals the number of 
stars in the system under study, which facilitates a correct proportion between 
various dynamical processes.
A third code of the same family was developed by \citet{FreitagBenz2001,
FreitagBenz2002} for studying dense galactic nuclei, featuring accurate treatment 
of loss-cone effects (including a timestep adjustment algorithm similar to that of 
Shapiro), and a model for physical collisions based on a large library of smooth 
particle hydrodynamics (SPH) simulations. 
Table~\ref{tab:compare_mc} compares the features of various Monte Carlo methods.

\begin{table*}
\begin{minipage}{17.7cm}
\caption{Comparison of Monte Carlo methods}  \label{tab:compare_mc}
\begin{tabular}{l p{3cm} p{4cm} l l l p{3.7cm}}
Name & Reference & relaxation treatment & timestep & 
1:1\footnote{One-to-one correspondence between particles and stars in the system}\!\!\! & 
BH\footnote{Massive black hole in the centre, loss-cone effects}\!\!\! & remarks\\
\hline
Princeton & \citet{SpitzerHart1971a, SpitzerThuan1972}\!\! & local dif.coefs.\ in velocity, 
  Maxwellian background $f(r,v)$ & $\propto \tdyn$ & no & no \\[0.55cm]
Cornell & \citet{MarchantShapiro1980} & dif.coef.\ in $E$, $L$, self-consistent background 
  $f(E)$ &   indiv., $\tdyn$ & no & yes & particle cloning\\[0.55cm]
H\'enon & \citet{Henon1971a} & local pairwise interaction, self-consistent bkgr. 
  $f(r,v_\|,v_\bot)$ & $\propto \trel$ & no & no \\[0.55cm]
& \citet{Stodolkiewicz1982} & H\'enon's & block, $\trel(r)$ & no & no & mass spectrum, disc shocks \\
& \citet{Stodolkiewicz1986} & & & & & binaries, stellar evolution \\[0.2cm]
& \citet{Giersz1998} & same & same & yes & no & 3-body scattering (analyt.) \\
\textsc{Mocca} & \citet{HypkiGiersz2013} & same & same & yes & no & 
  single/binary stellar evol., few-body scattering (num.) \\[0.2cm]
& \citet{JoshiRPZ2000} & same & $\propto\trel(\mathrm{center})$\! & yes & no & partially parallelized \\
\textsc{Cmc} & \citet{UmbreitFCR2012}, Pattabiraman+ (2013) & & (shared) & yes & yes & 
  few-body interaction, single/ binary stellar evol., GPU\\[0.2cm]
\textsc{Me(ssy)${}^2$} & \citet{FreitagBenz2002} & same & 
  indiv.$\propto\trel$ & no & yes & cloning, SPH physical collis.\\[0.2cm]
\Raga & this study & local dif.coef.\ in velocity, self-consistent background $f(E)$ & 
  indiv.$\propto \tdyn$ & no & yes & arbitrary geometry
\end{tabular}
\end{minipage}
\end{table*}

In parallel with the Monte Carlo codes, the approach based on direct integration of the 
Fokker--Planck equation using finite-difference schemes was developed by \citet{Cohn1979,
Cohn1980}, and later by \citet{Takahashi1993,Takahashi1995} and \citet{DrukierCLY1999} 
for spherical systems. However, it seems to be impractical to extend it beyond two-integral 
axisymmetric case \citep{Goodman1983,EinselSpurzem1999,FiestasSpurzem2010}, 
as the method relies on the explicit knowledge of integrals of motion. 
Another related method is the gaseous model, in which the relaxation is treated using 
a conductive approximation \citep{LouisSpurzem1991,AmaroSeoaneFS2004}, and which can 
be combined with Monte Carlo treatment of stellar binaries \citep{SpurzemGiersz1996}. 
This approach also was developed in the spherical case only.

With the advent of special-purpose hardware in 1990s, it became possible to perform 
direct \Nbody simulations of globular clusters with more than $10^5$ stars
\citep{Makino1996,BaumgardtME2004}, and a wide range of physics may be added to 
the dynamical evolution \citep[e.g.][]{Pelupessy2013}. These simulations are also 
not restricted to any particular geometry, but are very computationally expensive and, 
as we will show, still practically unsuitable for some classes of problems, in 
which collisional relaxation should be rather small compared to collisionless 
effects arising from non-spherical mass distribution. 
A number of more esoteric approaches have been proposed to combine the flexibility 
of collisional direct \Nbody simulations with 
Fokker--Planck \citep[e.g.][]{McMillanLightman1984}, 
spherical-harmonic expansion \citep[e.g.][]{HemsendorfSS2002}, 
self-similar dynamic renormalization \citep{SzellMK2005}, 
or tree-code \citep{McMillanAarseth1993,FujiiIFM2011,OshinoFM2011}, 
none of which apparently gained substantial popularity.

From the side of collisionless simulations of nearly-equilibrium systems, 
the most relevant for this study are the spherical-harmonic methods \citep[e.g.][]%
{Aarseth1967,CluttonBrock1973,vanAlbada1977,HernquistOstriker1992,MeironLHBS2014}. 
In this approach, the smooth potential of a stellar system is represented as a sum of
angular harmonics, with the radial variation of the expansion coefficients being 
either an explicit function of radius, or another sum over several basis functions. 
The coefficients of expansion are computed from the spatial distribution of particles, 
and the equations of motion of particles are governed by the gradients of this smooth 
potential. In all existing implementations, however, the timestep for particle motion, 
which is necessarily a small fraction of dynamical time, is also used for updating 
the coefficients of expansion, thereby imposing random fluctuations on them which 
effectively create numerical noise comparable to that of more direct \Nbody 
methods \citep{HernquistBarnes1990}. On the other hand, using some sort of temporal 
smoothing for the expansion coefficients \citep[proposed by][but not tried]
{HernquistOstriker1992} might help 
to reduce the unwanted relaxation considerably below the discreteness limit, 
while retaining the ability to follow the slow evolution of a non-spherical system.
A similar idea was recently used by \citet{BrockampBK2014} in the context of 
evolution of the population of globular clusters in the galaxy.

The idea to marry the benefits of collisionless expansion codes and collisional 
Monte Carlo approach has led us to a new formulation of the Monte Carlo method that 
avoids the restriction to spherical symmetry while retaining the ability to model 
the two-body relaxation rather faithfully. 
In essence, it is a successor to the Spitzer's variant of Monte Carlo method, 
with the orbits of test stars followed in real space in a smooth potential 
represented by a suitable expansion, and perturbations are applied to particle 
velocities in accordance with local diffusion coefficients. 
The new method is dubbed the \Raga code, which stands for ``Relaxation in Arbitrary 
Geometry'', but also alludes to slowly developing musical themes in the classical 
Indian tradition.
In addition to this method, we have also implemented a variant of spherical isotropic 
Fokker--Planck code, similar to that of \citet{Cohn1980}, and an orbit-averaged 
spherical isotropic Monte Carlo code, a simplified version of the method of
\citet{MarchantShapiro1980}, mainly for the purpose of testing the main code.
Below, we present a complete mathematical description and test simulations.

\section{Two-body relaxation}  \label{sec:relaxation}

In this section we review the standard two-body relaxation theory as used in our code, 
referring to \citet[][Chapter 5]{MerrittBook} for a more complete description.
As in most previous studies (the notable exception being Monte Carlo codes based on 
the H\'enon's approach), we consider scattering of test particles by an isotropic 
spherically symmetric population of background particles, described by the 
mass distribution function $f(\boldsymbol{x},\boldsymbol{v}) = f(E)$, where 
$E\equiv \Phi(\boldsymbol{x})+\boldsymbol{v}^2/2$ is the energy per unit mass. 
The scattering is described in terms of local (position-dependent) velocity drift and 
diffusion coefficients \citep[e.g.][Equations 5.23, 5.55]{MerrittBook}%
\footnote{Here, we make the usual Chandrasekhar's approximation by assuming that the 
drift (friction) term is determined only by background stars that move more slowly than 
the test star. One should keep in mind that in some situations this approximation 
breaks down \citep{AntoniniMerritt2012}.}:
\begin{subequations}  \label{eq:dc_vel}
\begin{flalign}
v\dvpar &= \textstyle -\left(1+\frac{m}{m_\star}\right) I_{1/2} \;,\\
\dvsqpar&= \textstyle \frac{2}{3} \left(I_0 + I_{3/2}\right) , \\
\dvsqper&= \textstyle \frac{2}{3} \left(2 I_0 + 3 I_{1/2} - I_{3/2}\right) ,
\end{flalign}
where $m$ and $m_\star$ are masses of the test and field stars, correspondingly, and 
\begin{flalign}
I_0     &\equiv \Gamma \int_E^0 dE'\,f(E') , \\
I_{n/2} &\equiv \Gamma \int_{\Phi(r)}^E dE'\,f(E') \left(\frac{E'-\Phi}{E-\Phi}\right)^{n/2} , \\
\Gamma  &\equiv 16\pi^2G^2m_\star\ln\Lambda = 16\pi^2 G^2 M_\mathrm{tot} \times 
(N_\star^{-1}\ln\Lambda) . \label{eq:perturbation_term}
\end{flalign}
\end{subequations}

These coefficients represent mean and mean-squared changes in velocity per unit time.
In the last equation, the term in brackets, or its appropriate generalization for 
a multimass case, is the only one that depends on $N_\star$ (for a given combination 
of $f,\Phi$). In the Monte Carlo code, we may assign the amplitude of perturbations 
at will, adjusting this term to a desired number of stars in the target system, which 
needs not be related to the number of particles in the simulation. In H\'enon's 
formulation, the particles were called ``superstars'', and their masses were a fixed 
multiple of actual stellar mass; contemporary codes usually have 1:1 correspondence 
between the particle and star mass, which facilitates the introduction of additional 
physical processes such as binary--single star scattering cross-section. 
In our approach, we do not require a fixed proportionality coefficient between 
the particle and star masses -- particles are just mass tracers and not actual stars, 
and the relaxation is determined by the smooth distribution function and not by 
discrete encounters.

In the rest of this section, we focus on the isotropic spherically symmetric case, 
which is used in the auxiliary codes, while the treatment of relaxation in the main 
\Raga code relies only on the velocity diffusion coefficients (\ref{eq:dc_vel}), 
with some secondary routines using the orbit-averaged energy diffusion coefficient.
Even in the spherical case, it is not necessary that the distribution of stars is 
isotropic: \citet{MarchantShapiro1980} worked in a two-dimensional $\{E,L\}$ phase-space, 
but retained the isotropic background approximation.

Local drift and diffusion coefficients in energy are
\begin{subequations}  \label{eq:dc_energy_local}
\begin{flalign}
\langle \Delta E \rangle   &= v\dvpar + \frac{1}{2}\dvsqpar + \frac{1}{2}\dvsqper 
 = I_0 - I_{1/2},\\
\langle \Delta E^2 \rangle &= v^2 \dvsqpar = \frac{2}{3} v^2 (I_0 + I_{3/2}).
\end{flalign}
\end{subequations}

The isotropic Fokker--Planck equation%
\footnote{It should be more appropriately called the generalized Landau equation, as it 
is a nonlinear integro-differential equation containing the unknown distribution function 
in the diffusion coefficients as well \citep{Chavanis2013}.}
describing the relaxation in energy has two forms.
The first is more convenient for Monte Carlo simulations, the second (flux-conservative) 
is more suitable for solving the Fokker--Planck equation on a grid.
Define $N(E) \equiv g(E) f(E)$ to be the mass density of stars per unit energy,
where the density of states 
\begin{equation}  \label{eq:ded}
g(E) \equiv 16\pi^2 \int_0^{\rmax(E)} dr\, r^2 v = 4\pi^2 \Lcirc^2(E) P(E) ,
\end{equation}
$\rmax(E)$ is the apocentre radius of a radial orbit with energy $E$ 
(so that $\Phi(\rmax(E))=E$),
$\Lcirc(E)$ is the angular momentum of a circular orbit with the given energy, 
and $P(E) \equiv 2\int_0^{\rmax} dr/v$ is the period of a radial orbit
(time needed to complete one oscillation in the radial direction).
Ignoring the time dependence of potential, the Fokker--Planck equation reads
\begin{equation}  \label{eq:fp1}
\frac{\d N(E)}{\d t} = -\frac{\d}{\d E} \left(N(E) \langle\Delta E\rangle_\mathrm{av} \right)
+ \frac{1}{2} \frac{\d^2}{\d E^2} \left(N(E) \langle\Delta E^2\rangle_\mathrm{av} \right) ,
\end{equation}
with $\langle \dots \rangle_\mathrm{av}$ being the averaged values of corresponding 
quantities over the phase volume accessible to the orbit. 
For the spherical case, these averages are given by
\begin{equation}
\langle \dots \rangle_\mathrm{av} = \frac{16\pi^2}{g(E)} 
\int_0^{\rmax(E)} dr\, r^2 v \, \langle \dots \rangle .
\end{equation}

The calculation of averaged coefficients for energy is made easier by introduction of 
a few auxiliary functions:
\begin{equation}  \label{eq:phasevol}
h(E) \equiv \frac{16\pi^2}{3} \int_0^{\rmax(E)} dr\, r^2 v^3 = 
  \int_{\Phi(0)}^E dE'\,g(E') \;,
\end{equation}
\begin{subequations}  \label{eq:K}
\begin{flalign}  
K_1(E) &\equiv  \int_E^0 dE'\,f(E') \;, \\
K_g(E) &\equiv  \int_{\Phi(0)}^E dE'\,f(E') g(E') \;, \\
K_h(E) &\equiv  \int_{\Phi(0)}^E dE'\,f(E') h(E') \;.
\end{flalign}
\end{subequations}

The function $K_g(E)$ measures the mass of stars having energy below $E$, 
while $K_h$ does the same for kinetic energy (up to a factor $3/2$).
These three functions, together with $h(E)$ and its derivative $g(E)$, 
can be tabulated for the given combination of $f(E)$ and $\Phi(r)$ 
and cheaply interpolated to obtain the drift and diffusion coefficients:
\begin{subequations}  \label{eq:dc_energy_av}
\begin{flalign}
\langle \Delta E \rangle_\mathrm{av}   &=  \Gamma [K_1(E) - K_g(E)/g(E) ],\\
\langle \Delta E^2 \rangle_\mathrm{av} &= 2\Gamma [K_1(E) h(E) + K_h(E) ]/g(E).
\end{flalign}
\end{subequations}

\section{The new Monte Carlo method}  \label{sec:raga}

Having reviewed the theory of two-body relaxation, we now describe the implementation 
of the new Monte Carlo code. In the present form, it is hardly suitable for realistic 
dynamical simulations of star clusters, lacking many sophistications found in other 
existing codes. We assume a population of identical single stars, neglect the dynamical 
influence of binaries and stellar evolution, and do not consider external tidal forces 
that would lead to escape of stars from the systems. Most of these ingredients are 
not difficult to add; the purpose of this paper is to show the feasibility and benefits 
of non-spherical dynamical Monte Carlo modelling.

The main advantage of position-dependent diffusion coefficients in velocity is that 
one may apply them to orbits of arbitrary shape, not restricted to spherical symmetry.
For a very general and flexible representation of the potential, we use two variants 
of spherical-harmonic expansions implemented in the publicly available \SMILE  
software \citep{Vasiliev2013}: basis-set and spline expansions. 
In both cases, the angular dependence of the potential is given by spherical harmonics, 
while for the radial part either a finite sum over a particular set of basis functions 
with adjustable coefficients is used, or the radial dependence of each spherical 
harmonic is represented with a spline function. 
This representation typically uses $10-20$ radial terms, and the order of the angular 
expansion $l_\mathrm{max}=4-6$ is sufficient for moderately flattened systems 
(with major to minor axis ratio $\lesssim 2$).
We refer to the appendix of the above paper for more details. 
Throughout this section, we denote the actual non-spherical potential in which 
the particles move as $\tilde\Phi(\boldsymbol r)\equiv \tilde\Phi(r,\theta,\phi)$, 
and its associated density as $\tilde\rho$, while the quantities from an equivalent 
spherical system, approximating the actual density profile (see below), are without tildes.

The evolution of the \Nbody system is followed through a series of ``episodes'' -- 
intervals of time $T_\mathrm e$ during which all particles move along their orbits 
independently from each other (thus the computation of orbits is trivially 
parallelized). At the end of an episode, the global state of the system (the potential 
and the diffusion coefficients) is updated using the orbits of particles during 
the episode: each orbit is sampled with $N_\mathrm{samp}$ points (position and 
velocity of the given particle at regular intervals of time). 
If $N_\mathrm{samp}\gg1$, this increases the effective number of particles used in 
recomputing the potential and distribution function, reducing the discreteness noise.

The motion of particles in the smooth potential of the entire system is computed 
using one of the ODE integrators from \SMILE: 
a standard eighth-order Runge--Kutta method \textsc{dop853} \citep{DOP853}, 
or several other methods from the \textsc{odeint} package \citep{odeint}. 
After each timestep, the perturbations to the velocity are computed as 
\begin{subequations}  \label{eq:vel_changes}
\begin{flalign}
\Delta v_\|   &= \dvpar \Delta t + \zeta_1 \sqrt{\dvsqpar \Delta t} \,,\\
\Delta v_\bot &= \zeta_2 \sqrt{\dvsqpar \Delta t} \,,
\end{flalign}
\end{subequations}
where $\zeta_1,\zeta_2$ are two independent random numbers with standard normal 
distribution, $\Delta t$ is the timestep adjusted internally in the ODE integrator, 
and the diffusion coefficients are given by (\ref{eq:dc_vel}).
While there are more sophisticated methods for dealing with stochastic differential 
equations \citep[e.g.][]{StochasticBook}, we used the simplest explicit order 0.5 
method for the stochastic part in combination with a high-order method for 
the deterministic part, which enables to follow the unperturbed trajectories with 
a great accuracy. We have checked that the choice of the integration method and 
the timestep criterion for the ODE integrator do not affect the statistical 
properties of accumulated changes in energy and angular momentum after a given 
interval of time $\gtrsim \tdyn$. Typically, the ODE integrator places several 
tens of timesteps and reaches the energy conservation error of better than 
$10^{-8}$ per $\tdyn$ for an unperturbed orbit. A similar approach (interleaving 
the evolution in the smooth field with the two-body perturbations) was used by 
\citet{WeinbergKatz2007} for studying the effect of noise on the behaviour of 
near-resonant orbits in spiral galaxies, and by \citet{JohnstonSH1999} for 
simulating the tidal mass-loss from galactic satellites (they used diffusion 
coefficient computed under the approximation of locally Maxwellian velocity 
distribution). 

The treatment of relaxation relies on the local diffusion coefficients 
(\ref{eq:dc_vel}) which are computed using an isotropic spherically symmetric 
equivalent of the system under study. This approximation (in particular, assumption 
of isotropy of velocities of background stars) is typical in the relaxation theory, 
however it could break down in a strongly non-spherical or anisotropic system.
A modification of the present approach could be adopted for a rotating stellar system, 
assuming isotropic velocity distribution in the corotating frame \citep{Goodman1983, 
EinselSpurzem1999}; we have not implemented it here.

At the beginning of the simulation, the equivalent spherical system is constructed 
by averaging the density profile $\tilde\rho(r,\theta,\phi)$ of the actual model 
over angles $\theta,\phi$, retaining only the radial dependence of mass profile $M(r)$. 
Then the associated spherically symmetric potential $\Phi(r)$ is computed, 
along with the isotropic distribution function $f(E)$ from the Eddington equation.
Later in the course of simulation, both the spherically symmetric mass profile 
which produces the associated potential, and the distribution function, are updated 
directly from the particle orbits (using points sampled during the episode).
Both the mass profile and the distribution function are constructed using 
a penalized spline smoothing approach \citep{Spline}, similar to the one employed 
in the \textsc{mkspherical} program from \SMILE. For the latter, we first compute 
$N(E)\,dE$ from the positions and velocities of sample points, using the spherically 
symmetric potential $\Phi(r)$, then smooth it, and finally $f(E)$ is obtained by
dividing the smoothed $N(E)$ by the density of states $g(E)$ (Equation~\ref{eq:ded}), 
again from the equivalent spherical system. 

After calculating the distribution function $f(E)$, we compute the functions 
$I_0$, $I_{n/2}$ that enter the definition of diffusion coefficients (\ref{eq:dc_vel}).
They depend on the energy of the test star $E$ and the (spherical) potential at 
the given position $\Phi(r)$, and we store the pre-computed functions on a grid in 
$\{E,\Phi\}$ space. In the course of orbit integration, the actual values of these 
functions are efficiently obtained from two-dimensional interpolation. 
For the given position and velocity of the particle in the actual (non-spherical) 
potential, the value of potential $\Phi$ and energy $E$ used as the arguments 
of these functions are taken from the spherically symmetric potential $\Phi(r)$ 
at the given position.
Thus, the actual potential $\tilde\Phi(r,\theta,\phi)$, responsible for the regular 
motion, and the diffusion coefficients for the stochastic perturbations, are computed 
using slightly different methods -- one for the actual system, the other for its 
spherical equivalent. The small incoherence amounts to the approximation of 
spherical isotropic scattering background, as used in most previous studies, 
and is believed not to cause substantial distortions to the dynamics. 
Note that the test stars themselves do not need to be isotropic in velocity -- 
this assumption is only used for the background stars.

In contrast with the H\'enon's formulation of Monte Carlo method, in which the 
energy is conserved by pairwise interactions (but not by the potential update; 
see \citet{Stodolkiewicz1982} for an amendment), the Spitzer and Shapiro's variants 
do not have this property intrinsically: each particle randomly walks in energy 
independently from others. To correct for this, at the end of the episode we compute 
the accumulated energy error and distribute it between particles, in proportion with 
their average diffusion coefficient $\sqrt{\langle \Delta E^2\rangle_\mathrm{av}}$ 
(\ref{eq:dc_energy_av}) during this episode. This is slightly different from 
the correction method employed by \citet{MarchantShapiro1980} and primarily applies 
the correction to those particles that have experienced the largest diffusion. 
We compensate the energy error by correcting the particle velocity at the end of 
the episode, changing its magnitude (but not direction) by a necessary amount. 

The true non-spherical potential used to compute particle motion is also updated 
at the end of an episode, using the same sampling points ($N_\mathrm{samp}$ per orbit)
collected during the episode. As already mentioned, $N_\mathrm{samp}\gg1$ reduces 
discreteness noise in the potential expansion coefficients; furthermore, if 
$T_\mathrm{e} \gg \tdyn(E)$ for most of the orbits in the system, each particle 
completes many periods during one episode, thus again smoothing out fluctuations. 
This is, in essence, the ``temporal smoothing'' proposed by 
\citet{HernquistOstriker1992} but apparently never used before.
(Note that there exist simulation methods that rely on ``orbit--orbit'', as opposed to 
``particle--particle'' interactions, which are used for stars on near-Keplerian orbits 
around a massive black hole \citep{ToumaTK2009,KocsisTremaine2014,HamersPZM2014}; 
these approaches hardly can be generalized for arbitrary potentials not dominated by 
a single point mass).

The energies of particles also need to be corrected after reinitialization of 
the potential, to account for the time dependence of the potential.
We adopt the method used by \citet{Stodolkiewicz1982} and \citet{Giersz1998}, 
which states that the energy correction for a given particle is 
\begin{equation}
\Delta \tilde E_i = \frac{1}{2}\left[
  \Delta\tilde\Phi(\boldsymbol r_{i,\mathrm{old}}) + 
  \Delta\tilde\Phi(\boldsymbol r_{i,\mathrm{new}}) \right],
\end{equation}
where $\boldsymbol r_{i,\mathrm{old}}$ and $\boldsymbol r_{i,\mathrm{new}}$ are 
particle positions at the beginning and end of the episode, and 
$\Delta\tilde\Phi(\boldsymbol r) \equiv 
\tilde\Phi_\mathrm{upd}(\boldsymbol r) - \tilde\Phi_\mathrm{old}(\boldsymbol r)$ 
is the difference between the updated potential and the old one (used during the episode). 
In these papers, this correction could be applied directly to the particle's energy, 
while in our case we are again forced to attribute it to the kinetic energy only. 
More specifically, the updated velocity after the correction is related to the ``new'' 
velocity (at the end of the episode but before the correction) by 
\begin{flalign}
\frac{v_{i,\mathrm{upd}}^2}2 + \tilde\Phi_\mathrm{upd}(\boldsymbol r_{i,\mathrm{new}}) &= 
\frac{v_{i,\mathrm{new}}^2}2 + \tilde\Phi_\mathrm{old}(\boldsymbol r_{i,\mathrm{new}}) +
\Delta \tilde E_i \;,  \nonumber\\
v_{i,\mathrm{upd}}^2-v_{i,\mathrm{new}}^2 &= 
\Delta\tilde\Phi(\boldsymbol r_{i,\mathrm{old}}) - 
\Delta\tilde\Phi(\boldsymbol r_{i,\mathrm{new}}) .
\end{flalign}

As shown by \citet{Stodolkiewicz1982}, this correction ensures the conservation of 
total energy of the system 
$\mathcal E \equiv \sum_i m_i\left[ v_i^2/2 + \tilde\Phi(r_i)/2 \right]$ in his case.
Indeed,
\begin{subequations}
\begin{flalign}
& \mathcal E_\mathrm{upd}-\mathcal E_\mathrm{old} = \nonumber\\ 
& \sum_i m_i \left[ 
  \frac{v_{i,\mathrm{upd}}^2}2 + \frac{\tilde\Phi_\mathrm{upd}(\boldsymbol r_{i,\mathrm{new}})}2 -
  \frac{v_{i,\mathrm{old}}^2}2 - \frac{\tilde\Phi_\mathrm{old}(\boldsymbol r_{i,\mathrm{old}})}2 
\right] = \nonumber\\ 
& \sum_i m_i \left[\frac{v_{i,\mathrm{new}}^2}2 + 
  \tilde\Phi_\mathrm{old}(\boldsymbol r_{i,\mathrm{new}}) \right] - \label{eq:sumEnew} \\
& \sum_i m_i \left[\frac{v_{i,\mathrm{old}}^2}2 + 
  \tilde\Phi_\mathrm{old}(\boldsymbol r_{i,\mathrm{old}}) \right] + \label{eq:sumEold} \\
& \sum_i m_i \frac{\tilde\Phi_\mathrm{upd}(\boldsymbol r_{i,\mathrm{old}}) - 
  \tilde\Phi_\mathrm{old}(\boldsymbol r_{i,\mathrm{new}}) } 2.  \label{eq:difPot}
\end{flalign}
\end{subequations}
The first two terms (\ref{eq:sumEnew}) and (\ref{eq:sumEold}) represent the sum of 
energies of all particles, which is conserved by the relaxation step followed by 
the cancellation of fluctuations described above. 
The last term (\ref{eq:difPot}) also should tend to zero in the continuum limit 
\citep[Eq.37]{Stodolkiewicz1982}. However, his proof is valid only if the updated 
potential is computed from the positions of particles at the end of the episode; 
if we use $N_\mathrm{samp}>1$ sampling points, this is no longer true. 
Therefore, we compute the last term explicitly, and cancel the total energy error 
$\mathcal E_\mathrm{upd}-\mathcal E_\mathrm{old}$ by distributing it between all 
particles.

In all these correction steps, we can only attribute the energy error to 
the kinetic energy by changing the magnitude of particle velocities. 
This might introduce some bias, as the energy excess/deficit is attributed entirely 
to the kinetic energy (and furthermore, if the energy needs to be subtracted from 
a particle happened to be around its turning point, the velocity may be too small 
to allow it, in which case it remains undercorrected), but it is the simplest 
practical way of cancelling the energy errors. 
We do not apply a similar correction to angular momentum fluctuations, as they 
remain small ($\sim N^{-1/2}$) and do not have a preferred sign (however, we only 
considered systems with zero total angular momentum and cannot be sure that there 
will be no secular drift in angular momentum if it was non-zero initially).

For the two auxiliary methods used for comparison -- spherical isotropic 
Fokker--Planck and Monte Carlo codes -- there is no need to follow orbits in space, 
only the evolution of distribution function $f(E)$ and its associated 
potential--density pair $\rho(r),\Phi(r)$. 
In the finite-difference Fokker--Planck scheme, $f(E)$ is sampled on a non-uniform 
grid in $E$, and a flux-conservative implicit \citet{ChangCooper1970} scheme is used
\citep[see][ for a comparison of numerical methods]{ParkPetrosian1996}.
In the Monte Carlo scheme, the distribution function is sampled by discrete 
particles with energies $E_i$, and during each episode, each particle performs 
one or more Monte Carlo steps with timestep $\Delta t$, according to 
\begin{equation}
\Delta E_i = \langle \Delta E \rangle_\mathrm{av} \Delta t 
  + \zeta \sqrt{\langle \Delta E^2 \rangle_\mathrm{av} \Delta t} \,,
\end{equation}
with $\zeta$ being a random number with standard normal distribution, and the diffusion 
coefficients given by (\ref{eq:dc_energy_av}). The timestep is assigned so that 
the expected change in energy does not exceed $\eta\,\mathrm{min}(|E|, E-\Phi(0))$, 
with the tolerance parameter $\eta \simeq 0.2$. When all particles have completed 
the episode, a new distribution function is computed in the same way as in the 
full \Raga code (i.e., using penalized spline smoothing).

The spherical potential is updated after the new distribution function has been 
computed, by using the following relation for the density:
\begin{equation}  \label{eq:density_from_df}
\rho(r) = 4\pi \int_{\Phi(r)}^0 dE\,f(E)\,\sqrt{2(E-\Phi(r))} \,,
\end{equation}
and then the Poisson equation for the potential. 
Followed by recomputation of the potential, the distribution function must be changed 
adiabatically, which is easiest to achieve by expressing it in terms of the phase 
volume $h$ (\ref{eq:phasevol}) instead of $E$, and then transforming back using the 
updated potential. As the equation (\ref{eq:density_from_df}) contains the unknown 
potential itself, it should be applied iteratively until convergence, while 
keeping $f(h(E))$ constant at each iteration while $E$ changes. In practice, 
we found that for the Fokker--Planck method one iteration is sufficient, provided 
that timestep for the update is small enough; for the spherical Monte Carlo code,
we perform several iterations to reduce the impact of fluctuations of potential 
at origin, where the number of particles is small. 
For a simulation of a deep collapse, the accumulated energy error is $\sim 1-2\%$.

\section{Tests}  \label{sec:tests}

\begin{figure*}
$$\includegraphics{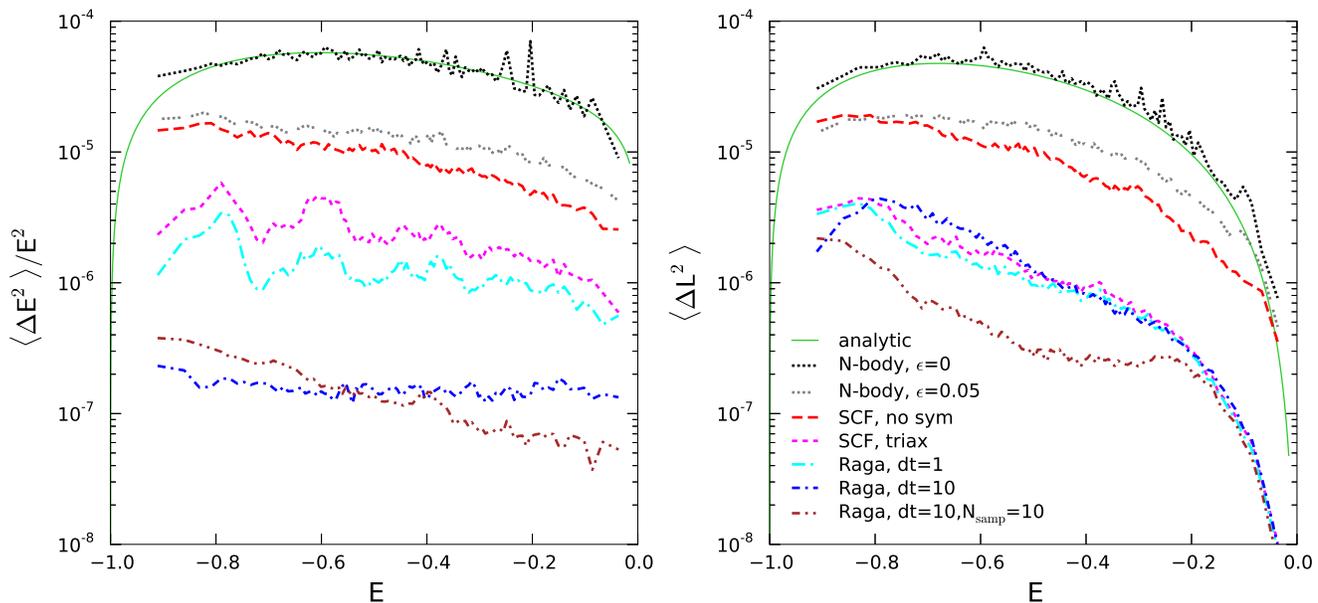}$$
\caption{ Comparison of energy (left) and angular momentum (right) diffusion rates 
as functions of energy, for various methods. 
The system under study is a spherical Plummer model with $N=10^5$ particles, 
evolved for $T=100$ time units (roughly $1/20$ of the half-mass relaxation time).
Plotted are mean squared changes in particle energy and angular momentum per unit 
time, averaged over $10^3$ particles in 100 energy bins. 
Energy diffusion coefficient is divided by squared energy for convenience.
The solid lines are the analytically computed diffusion coefficients, assuming 
the value of Coulomb logarithm $\ln\Lambda=9.3$. Dotted lines are the results of 
a direct \Nbody simulation, which agree well with the analytical predictions. 
Dashed lines are from SCF simulations with $n_\mathrm{max}=15$ radial and 
$l_\mathrm{max}=4$ angular coefficients: top is for the simulation retaining all 
coefficients, bottom is for the one with imposed triaxial symmetry (retaining only 
cosine coefficients with even $l,m$). The timestep for the SCF simulation is 
$1/32$, or $\sim 10^{-2}$ dynamical times at the centre. 
Other lines are for the Monte Carlo code with relaxation switched off, and 
using the same potential representation (Hernquist--Ostriker basis set with 
the same number of coefficients and imposed triaxial symmetry), 
but different settings for update interval:
1 time unit, 10 time units (longer than the dynamical time for most particles), 
and 10 time units with 10 sampling points for each particle during this interval.
Clearly, introduction of temporal softening (in terms of longer update interval) 
and increase of the number of sampling points decreases the fluctuations of 
the potential and reduces the diffusion of energy and angular momentum by as much 
as two orders of magnitude, with respect to the direct \Nbody simulation. 
A Monte Carlo simulation with relaxation included (not shown) produces 
the diffusion rate in very close agreement with the analytic predictions.
}  \label{fig:energydiff}
\end{figure*}

\begin{figure*}
$$\includegraphics{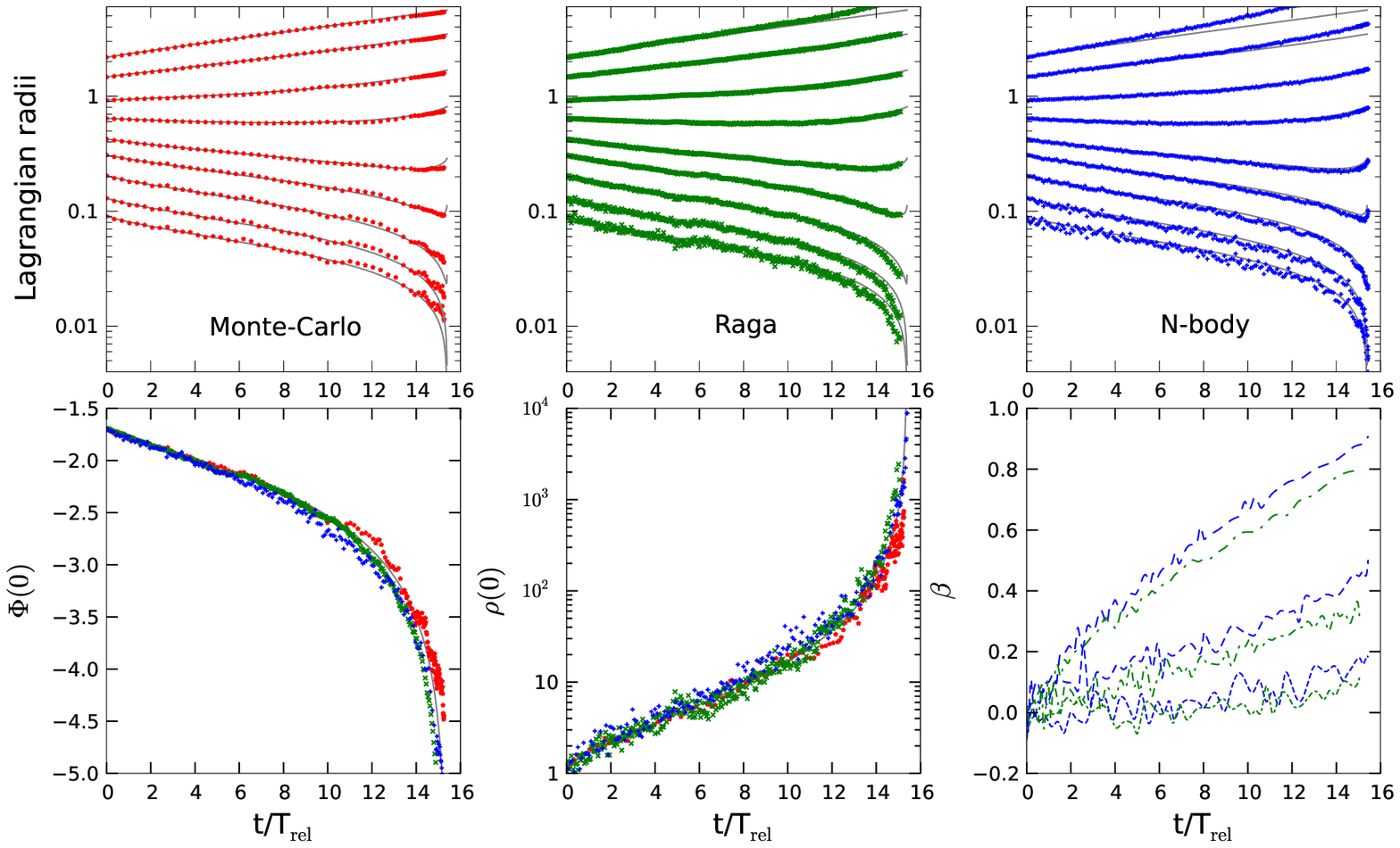}$$
\caption{ 
Comparison of time evolution of various quantities in a simulation of a Plummer model
undergoing core collapse, performed with the spherical Fokker--Planck (grey lines) 
and Monte Carlo (red stars) methods, the \Raga code (green diagonal crosses) 
and a direct \Nbody simulation (blue horizontal crosses). 
Time is expressed in units of half-mass relaxation time 
($\trelhm\equiv 0.093\,N/\ln\Lambda$), with the adopted value $\ln\Lambda=6.5$ for 
the \Nbody simulation being lower than the usually employed $\ln\Lambda=\ln0.11\,N
\approx 7.5$, thus bringing the collapse time in this particular simulation into 
better agreement with the Fokker--Planck simulation. \protect\\
Top row: Lagrangian radii containing the fraction of mass equal to 
0.0035, 0.01, 0.035, 0.1, 0.2, 0.4, 0.6, 0.8, 0.9 (from bottom to top).
Bottom left: the value of potential in the centre 
(starting from $-16/(3\pi)\approx -1.7$).
Bottom centre: central density. 
Bottom right: velocity anisotropy parameter $\beta\equiv 1-\frac{\sigma_t^2}{2\sigma_r^2}$
(where $\sigma_t$ and $\sigma_r$ are the transversal and radial velocity dispersions),
evaluated in three shells, enclosed by Lagrangian radii containing 
45 to 50, 70 to 75, and 90 to 95\% of total mass (from bottom to top).
Dashed lines are from \Nbody simulation and dot--dashed -- from the \Raga simulation 
(the other two methods assume isotropic velocity).
} \label{fig:core_collapse_comparison}
\end{figure*}

In this section, we describe several test problems for the new Monte Carlo method.
First, we demonstrate that temporal smoothing does help to reduce energy exchange 
between particles due to fluctuations of the potential to a negligible level, 
compared with the typical two-body relaxation rates. 
Then, we perform two standard tests: the core collapse of a Plummer sphere, 
and the growth of a Bahcall--Wolf cusp around a massive black hole. 
Finally, we consider the shape evolution of a triaxial model with a black hole.

\subsection{Temporal smoothing test}  \label{sec:test_temporal_smoothing}

In this test we consider the relaxation rate of a spherical Plummer model, 
evolved with different methods: \Nbody simulation with a direct-summation code, 
self-consistent field (SCF) method, and the \Raga code with relaxation turned off.
The goal is to demonstrate that temporal smoothing does substantially reduce 
the energy and angular momentum relaxation rate, compared to more direct simulation 
methods. We take an $N=10^5$ Plummer model with total mass and scale radius both 
equal to 1 \Nbody units, and evolve it for $T=100$ time units, or roughly $1/20$ 
of the half-mass relaxation time. 
To measure the relaxation rate, we record the changes in energy and angular momentum 
of individual particles, average them over particles in each of 100 bins sorted 
in energy, and fit a linear regression to the squared difference between initial 
and current values of $E$ and $L$ as functions of time (see \citet{Theuns1996} 
for a somewhat different method of estimating the relaxation rate). 
The coefficient of this regression represents the diffusion coefficient 
$\langle \Delta E^2\rangle_\mathrm{av}$ (\ref{eq:dc_energy_av}) and a similarly 
computed coefficient for $L$. We have checked that the growth of $\Delta E^2$ and 
$\Delta L^2$ is indeed close to linear in time, with occasional fluctuations.

For the conventional \Nbody simulation we use the GPU-accelerated direct-summation 
code \phiGRAPE \citep{HarfstGMSPB2007} with the \textsc{sapporo} library 
\citep{GaburovHP2009}. 
Figure~\ref{fig:energydiff} demonstrates that the theoretically computed diffusion 
rates agree very well with the measured values from the direct \Nbody simulation 
without softening, using the standard value of the Coulomb logarithm 
$\ln\Lambda=\ln 0.11\,N\approx 9.3$ \citep{GierszHeggie1994}. 
In collisionless simulations, softening is used to reduce the graininess of the 
potential; we have run another simulation with $\epsilon=0.05$, which is close to 
the optimal value for this $N$ \citep{Merritt1996,AthanassoulaBLM1998} and reduces 
the relaxation rate by a factor of few \citep{Theis1998}.
The other, ``indirect'' \Nbody simulation method that we used was the SCF method 
of \citet{HernquistOstriker1992}, employed in two regimes: 
in the first case we used all expansion coefficients ($n_\mathrm{max}=15$ radial and 
$l_\mathrm{max}=4$ angular terms), in the second -- retained only the non-zero terms 
for a triaxially symmetric model (that is, cosine terms with even $l$ and $m$).
Figure~\ref{fig:energydiff} shows that the SCF method demonstrates a several times 
lower rate of diffusion than a direct \Nbody simulation, when using all coefficients, 
and a further factor of few lower rate for a model with imposed triaxial symmetry. 
This is not unexpected, given that the potential in the SCF method is fairly smooth, 
but the reduction of relaxation rate is limited by the fluctuations in the potential 
arising from frequent updates in the coefficients, as the update interval is equal 
to the timestep of equations of motion (taken to be a small fraction ($\sim 10^{-2}$)
of the dynamical time in centre), and is comparable to the reduction due to softening 
in a direct \Nbody simulation \citep{HernquistBarnes1990}. It can further be improved 
by a factor of few by using a carefully constructed basis set \citep{Weinberg1996}.

On the other hand, if we allow for less frequent updates in the potential while 
retaining the high accuracy in integrating the equations of motion, then the relaxation 
rate may be reduced even further, as shown by the simulations of \Raga code with 
longer update intervals (we checked that running it with the same timestep as the SCF 
code produced identical results to the latter). 
Increasing the number of sampling points $N_\mathrm{samp}$ for each particle per episode 
reduces the fluctuations even further. 
Overall, for this model we attained a factor of hundred reduction in the relaxation 
rate, limited only by the update frequency: if the system needs to be simulated for 
a time substantially shorter than its relaxation time (or the time for any other 
effect to change its structure significantly), then the potential update may be 
switched off altogether, entirely eliminating this source of unwanted fluctuations. 
On the other hand, the necessary level of relaxation is readily restored by adding
the stochastic two-body perturbation term to the equations of motion. 
We have checked that this produced essentially the same total relaxation as the 
direct \Nbody simulation, if the amplitude of perturbation term in Equation~%
(\ref{eq:perturbation_term}) was assigned accordingly, using the same values of $N$ 
and $\Lambda$.

\subsection{Core collapse test}  \label{sec:test_core_collapse}

\begin{figure*}
$$\includegraphics{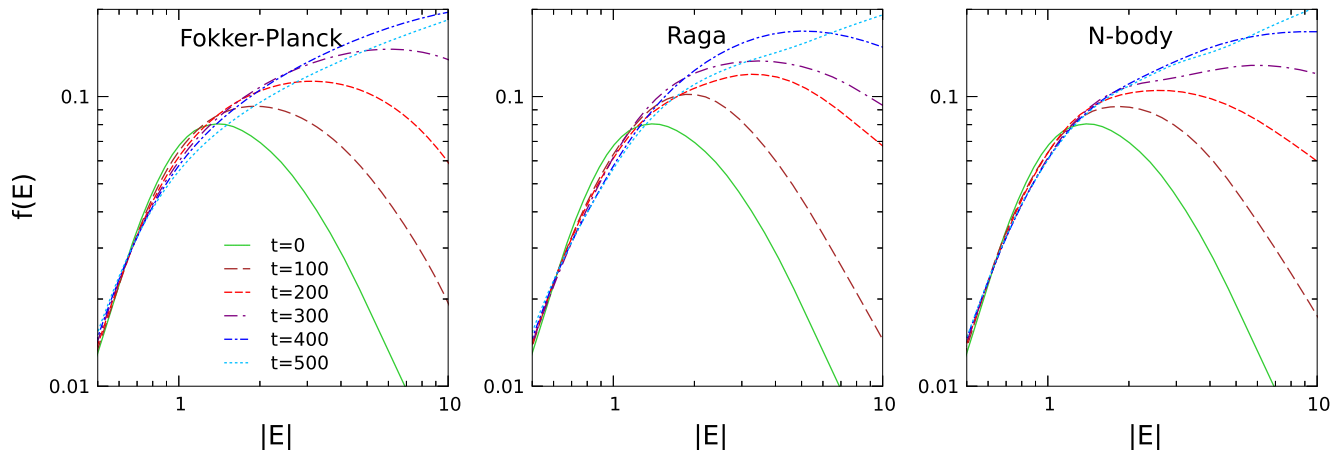}$$
\caption{ 
Evolution of distribution function in a system with a central black hole: 
left panel -- Fokker--Planck, centre -- \Raga, right -- \Nbody simulation.
The system is a $N=32000$ realization of a spherical $\gamma=1$ Dehnen model with 
the black hole mass $M_\bullet=0.1$ of the total mass in stars. 
The Bahcall--Wolf cusp with the density profile $\rho\propto r^{-7/4}$, corresponding 
to the distribution function $f(E) \propto |E|^{1/4}$, develops after roughly 400 
time units, or roughly $0.2\,\trel(\rh)$ (or one $\trel$ at a radius $\sim 0.2\rh$, 
where the relaxation time is shortest initially). We scaled the relaxation rate 
using the value of Coulomb logarithm $\ln\Lambda = \ln M_\bullet/m_\star \approx 8$.
Different codes show a rather good agreement in the evolution of distribution function 
towards the steady-state solution.
} \label{fig:bahcall_wolf}
\end{figure*}

Self-gravitating systems are known to have negative specific heat and exhibit 
the phenomenon of core collapse \citep[Chapter 18]{HeggieHutBook}. 
The easiest and probably most studied example is that of a Plummer sphere composed 
of equal-mass particles, for which various studies based on isotropic Fokker--Planck 
method have found the core collapse time to be $\approx 15$ initial half-mass relaxation 
times $\trelhm$ \citep[e.g.][]{SpitzerShull1975,Cohn1980,Takahashi1993,Quinlan1996}.
As discussed in the latter paper, a constant value of Coulomb logarithm overestimates 
the relaxation rate in the centre at later stages of core collapse, as the effective 
number of stars in the core decreases; anisotropic models also tend to have longer 
collapse times \citep[e.g.][]{Takahashi1995}. 

For this test, we set up an $N=16384$ Plummer model in the virial units (with 
the scale radius set to $3\pi/16$).
In the calibration \Nbody run, performed by the same code \phiGRAPE, 
the moment of collapse corresponds to $T_\mathrm{coll}\approx 3630$ time units, 
in agreement with other \Nbody studies \citep{Makino1996,BaumgardtHHM2003}.
With the standard choice of Coulomb logarithm $\ln\Lambda\approx 7.5$ this 
corresponds to $17.8\,\trelhm$; for the purpose of comparison with the isotropic 
Fokker--Planck and Monte Carlo simulations we have used a smaller value 
$\ln\Lambda=6.5$, which brings the collapse time in this particular simulation 
into better agreement with other methods.

Next we have run spherical isotropic Fokker--Planck and Monte Carlo codes, as well 
as the full \Raga code.
The Fokker--Planck simulation was taken as reference, with the time until core 
collapse being $15.4\,\trelhm$, in excellent agreement with other studies. 
Figure~\ref{fig:core_collapse_comparison} shows the evolution of various quantities 
(Lagrangian radii, central density and potential, and velocity anisotropy) 
in different simulations. Overall, the agreement between various methods is fairly 
good, at least until the central density increases by a factor of $10^3$; closer 
to the time of collapse, we do not expect either method to be reliable without taking 
into account the binary formation and heating and other phenomena beyond two-body 
relaxation. By the end of the simulation, the accumulated energy error was around 2\%.

\subsection{Bahcall--Wolf cusp growth test}  \label{sec:test_bahcall_wolf}

The density profile around a point mass (massive black hole) has a steady-state 
power-law solution of the Fokker--Planck equation, known as the \citet{BahcallWolf1976}
cusp: $\rho(r) \propto r^{-7/4}$ for a single-component star cluster.
Dynamical models starting from different initial conditions tend to develop 
the cusp at radii smaller than $r\sim 0.2\rh$, where the influence radius of 
the black hole $\rh$ contains the mass in stars equal to twice the black hole mass.
This has been observed both in Fokker--Planck models \citep[e.g.][]{CohnKulsrud1978,
Merritt2009} and in \Nbody simulations \citep{PretoMS2004,BaumgardtME2004,MerrittSzell2006}. 
As the relaxation time in the Newtonian potential of the central point mass 
is proportional to $\rho^{-1}(r)\,r^{-3/2}$, if the initial density profile was 
shallower than $r^{-3/2}$, then the cusp grows from outside in.

We have set up a spherical \citet{Dehnen1993} model with $\gamma=1$, $N=32000$ and 
a central black hole with mass $M_\bullet=0.1$ of the total mass in stars, drawing 
particle positions and velocities from a self-consistent isotropic distribution function 
\citep[e.g.][]{TremaineEtAl1994}, computed numerically from the Eddington's formula, 
using the \textsc{mkspherical} program from \textsc{smile}.
Then, we evolved the model until it developed the steady-state Bahcall--Wolf profile. 
The \Nbody simulation used a version of code with chain regularization 
\citep[{\phiGRAPE}ch,][]{HarfstGMM2008}. 
In the Fokker--Planck model we adopted a zero-flux boundary condition at the black hole.
Figure~\ref{fig:bahcall_wolf} shows the gradual evolution of the distribution 
function towards the $|E|^{1/4}$ solution. The agreement between Fokker--Planck, 
Monte Carlo and \Nbody simulations is again quite good.

\subsection{Shape evolution test}  \label{sec:test_shape_evolution}
\begin{figure*}
$$\includegraphics{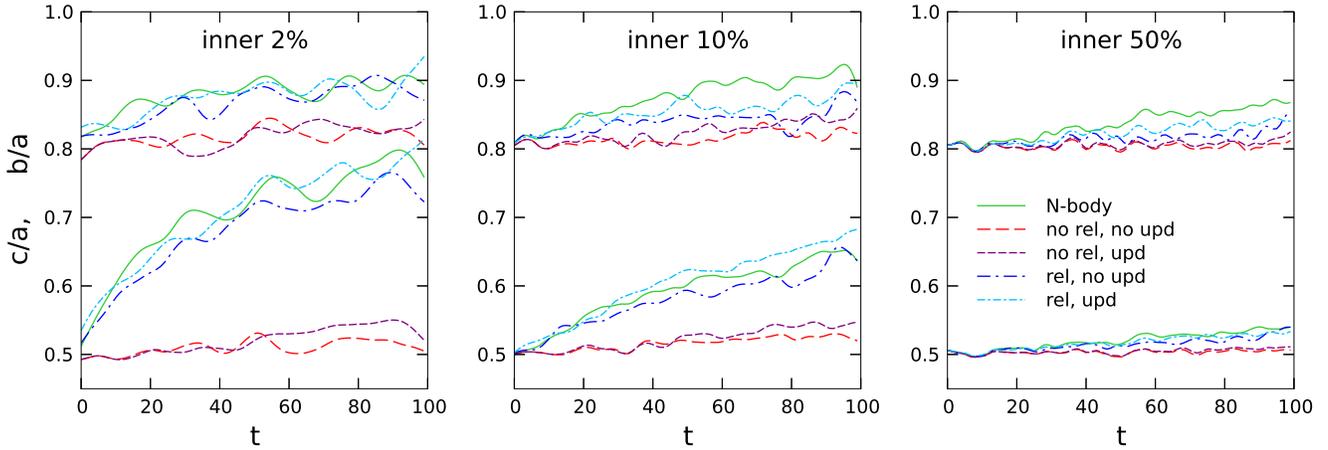}$$
\caption{ 
Evolution of shape of a triaxial $\gamma=1$ Dehnen model with a central black hole 
of mass $M_\bullet=0.01$ of the total model mass.
Three panels display the axis ratios ($b/a$ -- intermediate to major and $c/a$ -- 
minor to major) evaluated at a radius containing 2, 10 and 50\% of total mass 
(excluding the black hole). 
Solid lines are for the \Nbody simulation ($N=10^5$), other lines are for the 
Monte Carlo simulations with or without relaxation and with or without potential 
update (coefficients of potential expansion recomputed every 10 time units). 
Clearly, relaxation has a dramatic effect on the evolution of shape towards more 
sphericity in the central parts of the model, with the rate of evolution similar 
to that seen in the \Nbody simulation. The potential update does not seem to have 
a significant impact; if anything, it accelerates the evolution slightly.
} \label{fig:shape_evolution}
\end{figure*}

Up to now we have considered spherical systems, to facilitate comparison between 
various methods. We now turn to the unique feature of \Raga code, namely the ability 
to simulate systems of arbitrary geometry. For this test, we take a triaxial 
$\gamma=1$ Dehnen model with axis ratio of $1:0.8:0.5$ and a central black hole 
with mass $M_\bullet=0.01$ of the total model mass. This was one of the test models 
for the \SMILE code for \citet{Schwarzschild1979} modelling described in 
\citet[][Section 7.1]{Vasiliev2013}. 
The simulation was conducted with the {\phiGRAPE}ch code using $N=10^5$ 
particles. 
It was found that the model evolved towards a somewhat less flattened 
and less triaxial shape over the timescale of simulation (100 time units). 
Such evolution is not unexpected in light of previous studies 
\citep[e.g.][]{GerhardBinney1985,MerrittQuinlan1998}, although later papers suggested 
that the evolution may be not as rapid as found earlier \citep{HolleyMSHN2002,
PoonMerritt2004}. The driving force behind this shape evolution is thought to be 
the scattering of chaotic orbits by the central point mass: this would let them 
more uniformly populate the equipotential surface, which is typically rounder than 
the equidensity surface. However, the diffusion of chaotic orbits may be greatly 
facilitated by the graininess of potential \citep{PogorelovKandrup1999,KandrupPS2000}, 
and very little has been explored on this topic.

We performed simulations of the same system as studied in \citet{Vasiliev2013} with 
the Monte Carlo code, in several regimes, using a combination of two options: 
(i) without two-body relaxation or with the stochastic perturbation equivalent to 
the relaxation rate of an $N=10^5$ system, and (ii) using a fixed initial potential, 
or updating the potential every 10 time units (for a total simulation time of 100 
time units). 
We used the iterative method E1 of \citet{ZempGGK2011} for computing the axis ratios 
of our models as functions of radius (the same method was used in the previous paper). 

Figure~\ref{fig:shape_evolution} shows the evolution of shape for our four 
runs, together with the one from the \Nbody simulation.
Clearly, in the absence of relaxation the shape does not substantially evolve, 
regardless of whether we update the potential or keep it fixed. 
On the other hand, inclusion of relaxation dramatically accelerates the shape 
evolution in the central parts of the model, bringing it in good agreement with 
the results of \Nbody simulation.
This experiment suggests that the evolution of shape can be at least partially 
attributed to the discreteness noise which accelerates the chaotic diffusion. 
The substantial reduction of unwanted collisional relaxation offered by the presented 
Monte Carlo scheme offers new avenues in exploring the interplay between discreteness 
and chaos, enabling a more robust study of chaotic diffusion and its effect on 
the galaxy shape \citep[e.g.][]{VasilievAthanassoula2012}.

\section{Conclusions}  \label{sec:conclusions}

We have reviewed the existing methods for simulating the evolution of stellar systems 
driven by the two-body relaxation, and proposed a new variant of Monte Carlo method 
suitable for studying non-spherical systems. It combines the flexible representation 
of the smooth average potential in terms of spherical-harmonic expansion 
\citep[similar to the SCF method of][]{HernquistOstriker1992} with the Spitzer's 
approach to the description of two-body relaxation in terms of local (position-dependent) 
velocity diffusion coefficients. 
The orbits of particles are thus evolved on a dynamical timescale, with the two-body 
interaction between them mediated by the diffusion coefficients computed from a smooth, 
nearly-stationary distribution function (in a manner similar to the Shapiro's variant 
of the Monte Carlo method, but without orbit-averaging). 
We have shown that the method reproduces some standard evolutionary models, and has 
a substantially reduced artificial relaxation rate (related to random fluctuations 
in the potential expansion coefficients) compared to the SCF method.
The wall-clock computation time of the Monte Carlo code was within one hour for 
all simulations discussed in this paper (using a typical multi-core desktop), 
while some of the \Nbody simulations took a few days using high-end GPUs.
The \Raga code is made publicly available at \url{http://td.lpi.ru/~eugvas/raga/}; 
additionally, its inclusion into the \textsc{AMUSE} framework \citep{PortegiesZwart2013}
is underway.

In the present implementation, the Monte Carlo method has a number of limitations, 
most of which are not fundamental:

\begin{itemize}
  \item The Monte Carlo method (in this and other variants, with the possible exception 
of Spitzer's original formulation) is not suitable for systems which are not in 
dynamical equilibrium.
  \item The fluctuations in velocities (and, hence, energies) of particles are simulated 
independently from each other, which means that at the end of the Monte Carlo episode 
the total energy has, in general, a non-zero accumulated deviation. It is corrected 
by distributing this energy error between all particles, in proportion to their 
time-averaged energy diffusion coefficient, but the correction is applied to the 
magnitudes of velocity only. This could in principle bias the dynamics somewhat, but 
at least avoids much larger errors which occur without any such correction.
The total angular momentum of the system is not conserved, but its fluctuations due to 
discreteness noise are rather small for a reasonable particle number.
  \item The calculations assumed that all stars have the same mass. This is quite easy 
to generalize, by allowing each simulation particle to carry a given ``token'' dynamical 
mass (which enters the expression (\ref{eq:dc_vel}a) for the drift coefficient), 
and this mass needs not be related to the actual amount of mass that this particle 
contributes to the total potential. In other words, we generalize H\'enon's concept 
of ``superstars'' by completely separating the notions of dynamical and tracer mass.
Likewise, stellar evolution may be accounted for by allowing this token mass 
to change with time. 
We note that for all simulations in this paper, we scaled the diffusion coefficients 
in such a way as to model a system with the number of stars $N$ being the same as 
the number of particles in the model, but this was done only to facilitate comparison 
with \Nbody simulations and is not a restriction of the code.
  \item We ignored primordial and dynamically formed binaries and their contribution 
to the energy budget of the system, and did not consider the process of escape of stars. 
This could be implemented in a similar way to other state-of-the-art codes
\citep[e.g.][]{FregeauGJR2003, GierszSpurzem2003}.
  \item Stellar collisions in dense galactic nuclei may be accounted for by a scheme 
similar to \citet{DuncanShapiro1983} and \citet{FreitagBenz2002}.
  \item The discrete nature of mass tracers makes it difficult to simulate a system 
with high density contrast without resorting to mass refinement schemes. Fortunately, 
in our implementation, the mass of a particle can be set in an arbitrary way, for 
instance, creating initial conditions with higher mass resolution where necessary 
\citep[e.g.][]{ZempMSCM2008,ZhangMagorrian2008}. However, if the evolution time is 
substantially longer than the central relaxation time, particles will tend to mix 
in energy, erasing the effect of mass refinement. To combat this, an adaptive 
``creation--annihilation scheme'', such as that employed by \citet{ShapiroMarchant1978}
and \citet{FreitagBenz2002}, could be added to the algorithm. However, this mixing would 
also presumably drive the system towards spherical symmetry, so that the benefits of the 
arbitrary-geometry code would be irrelevant; for systems with longer relaxation times 
(such as galactic nuclei) the initial mass refinement should suffice.
  \item The diffusion coefficients are computed under the approximation of 
a spherically symmetric isotropic distribution function of background stars. 
This is perhaps the most fundamental limitation, and it means that we may reliably 
simulate only systems that are not too flattened and not too far from isotropy.
It is known that in stellar systems that are at least partially rotationally supported, 
the two-body relaxation proceeds faster as the velocity dispersion is lower 
\citep[e.g.][]{Goodman1983, Sellwood2013, SubrHaas2014}. However, it is possible to adapt 
the computation of diffusion coefficients for a distribution function that is isotropic 
in the rotating frame \citep{EinselSpurzem1999}.
  \item Similarly, we did not take into account the processes that are not described 
by standard two-body relaxation theory, such as resonant relaxation in the vicinity 
of a massive black hole \citep{RauchTremaine1996} or non-Gaussian character of energy 
diffusion at $t\ll \trel$ \citep{BarOrKA2013}. The proper account of these processes 
is hindered by the fact that they are not simply described by uncorrelated random walk, 
and require more sophisticated statistical models \citep[e.g.][]{MadiganHL2011}. 
The error introduced by neglecting these effects depends on the question being addressed.
For instance, the total rate of capture of stars by a massive black hole is rather 
weakly influenced by resonant relaxation, as shown by \citet{HopmanAlexander2006} as 
well as by our own comparison of direct \Nbody simulations with the Fokker--Planck models
\citep{VasilievMerritt2013}, because the bulk of captured stars come from larger energies
that those for which the resonant relaxation is effective. On the other hand, it surely 
is important for stars very close to the black hole, as are relativistic effects 
\citep[e.g.][]{MerrittAMW2011}, also ignored in this study.
\end{itemize}

The possibility of simulating collisional relaxation for stellar systems with arbitrary 
shape opens up a number of opportunities, especially for studies of elliptical galaxies 
and galactic nuclei which are otherwise inaccessible to direct \Nbody simulations 
with present-day computers:
\begin{itemize}
  \item Noise is known to enhance the efficiency of chaotic diffusion \citep[e.g.][]
{KandrupPS2000}, especially in systems with a rich population of sticky chaotic orbits 
\citep{HabibKM1997}, such as triaxial Dehnen models \citep{ValluriMerritt1998}, 
and has been proposed to improve the phase-space coverage of chaotic orbits in the 
construction of Schwarzschild models \citep{SiopisKandrup2000}.
However, very little is known of the implications of noise for the secular evolution of 
triaxial galaxies which may -- or may not, depending on their orbital structure -- evolve 
noticeably away from triaxiality during the Hubble time \citep{VasilievAthanassoula2012}.
  \item Non-spherical galactic nuclei have been proposed as a way to increase the rate 
of star captures by a supermassive black hole \citep{NormanSilk1983,GerhardBinney1985,
MerrittPoon2004,HolleySigurdsson2006}, if the triaxiality can persist for the Hubble time.
On the other hand, scattering of chaotic orbits by the black hole might destroy or 
reduce the triaxiality \citep{MerrittQuinlan1998,HolleyMSHN2002}, and the collisional 
relaxation increases the rate of diffusion of stars into the black hole even in the 
axisymmetric potential \citep{MagorrianTremaine1999,VasilievMerritt2013}. 
The evolution of non-spherical black hole nuclei, including the loss of stars into 
the black hole and changes in the galaxy shape, is difficult to follow by conventional 
\Nbody simulations because of very low relaxation rates in actual galaxies, compared to 
what can be achieved in the direct simulations. This topic is explored with the new 
Monte Carlo method in a separate paper \citep{Vasiliev2014}.
  \item Likewise, the dynamics of binary supermassive black holes is substantially 
changed in a non-spherical system \citep{BerczikMSB2006,PretoBBS2011,KhanJM2011,
KhanHBJ2013}, although an accurate treatment of collisional relaxation in the 
non-spherical case is even more difficult for a binary black hole 
\citep{VasilievAM2014}.
  \item After implementing mass-dependent velocity drift coefficient, it becomes possible 
to study dynamical friction of not too massive objects (heavier than field stars, but 
much lighter than the total mass of the model) in non-spherical galaxies 
\citep[e.g.][]{Binney1977,PesceCV1992,CoraVM2001,VicariCM2007} in a more self-consistent way 
(including possible feedback on the galaxy shape). A related idea was recently explored 
by \citet{BrockampBK2014}, although their generic machinery of basis-set expansion 
was only applied for the spherical case.
Likewise, collisional evolution and mass segregation in galactic nuclei has been mostly 
studied in the spherical case \citep{FreitagAK2006,AlexanderHopman2009,Merritt2010}; 
only a few studies have considered non-spherical nuclei, resulting, for instance, 
from galactic mergers \citep[e.g.][]{GualandrisMerritt2012} or globular cluster inspirals 
\citep[e.g.][]{Antonini2014}. 
  \item Accurate treatment of escape of stars from globular clusters in a realistic tidal 
field of a galaxy is non-trivial \citep{FukushigeHeggie2000}, and it is quite tricky to 
implement it in a spherical Monte Carlo code \citep{GierszHHH2013, SollimaMastrobuono2014}. 
More generally, non-spherical globular clusters may present other interesting phenomena 
to study \citep[e.g.][]{CarpinteroMW1999}.
  \item By modifying the expressions for diffusion coefficients using a suitable 
definition of background distribution function of stars (for instance, lifting 
the assumption of isotropy), it will be possible to study rotating clusters \citep[cf.]
[who used axisymmetric Fokker--Planck models]{EinselSpurzem1999,FiestasSpurzem2010}.
\end{itemize}

I warmly thank Douglas Heggie and Mirek Giersz for detailed comments on the early version 
of the manuscript, and am grateful to the anonymous referee for helpful remarks that 
improved the presentation.
This work was partly supported by the National Aeronautics and Space Administration 
under grant no. NNX13AG92G.

\label{lastpage}

\end{document}